%% file: main.tex
\documentclass[conference,compsoc]{IEEEtran}
\ifCLASSOPTIONcompsoc
  \usepackage[nocompress]{cite}
\else
  \usepackage{cite}
\fi
\newcommand*{\field}[1]{\mathbb{#1}}%

\ifCLASSINFOpdf
  \usepackage[pdftex]{graphicx}
\else
\fi

\usepackage{amsmath}

\usepackage[ruled, linesnumbered,noend]{algorithm2e}

\usepackage{array}

\ifCLASSOPTIONcompsoc
  \usepackage[caption=false,font=footnotesize,labelfont=sf,textfont=sf]{subfig}
\else
  \usepackage[caption=false,font=footnotesize]{subfig}
\fi
\usepackage{fixltx2e}
\usepackage{float}

\usepackage[multiple]{footmisc}
\usepackage{url}
\usepackage{esvect}
\usepackage{textcomp}
\usepackage{diagbox}
\usepackage{tabularx}
\usepackage{dcolumn}
\usepackage{booktabs}
\usepackage[english]{babel}
\usepackage[dvipsnames]{xcolor}
\usepackage[utf8]{inputenc}
\usepackage[T1]{fontenc}
\usepackage{latexsym}
\usepackage{amssymb}
\usepackage{listings}
\usepackage{caption}
\usepackage{subfig}

\usepackage{longtable}
\usepackage{rotating}
\usepackage{array}
\usepackage{multicol}
\usepackage{multirow}

\makeatletter
\newcommand{\removelatexerror}{\let\@latex@error\@gobble}
\makeatother

\makeatletter
\newcommand*{\rom}[1]{\expandafter\@slowromancap\romannumeral #1@}
\makeatother

\lstdefinelanguage{code}{
  morekeywords={let,in,def,aspect,before,after,pointcut,public,privileged,protected,declare,parents, call,target,implements,throw,new,for,class,forAll,exists,Boolean,return,break,executeCaller,true,
  false,pre,if,then,else,endif,String,join,select,from,where,with,create,temporary,table,and,
  or,as,on,case,when,end,union,iterate,intersection,symmetricDifference,self,includes,function, Text,Set,boolean, show,insert, into,div,mod,Integer, by, having, on, not, var, while, continue,switch,execute,abort,delete,eval,this,input, output,process},
  basicstyle=\footnotesize\usefont{T1}{pcr}{m}{n}\selectfont,
  keywordstyle=\footnotesize\usefont{T1}{pcr}{b}{n}\selectfont,
  identifierstyle=\footnotesize\usefont{T1}{pcr}{m}{n}\selectfont,
  commentstyle=,
  stringstyle=\footnotesize\usefont{T1}{pcr}{m}{n}\selectfont,
  numberstyle=\footnotesize,
  tabsize=2,
  frame=lines,
  upquote=true,
  xleftmargin=10pt
}

\usepackage{cite}

\begin{document}
%

\bstctlcite{IEEEexample:BSTcontrol}

\title{Hybrid-IoT: Hybrid Blockchain Architecture for Internet of Things - PoW Sub-blockchains}


\author{
    \IEEEauthorblockN{Gokhan Sagirlar\IEEEauthorrefmark{1}, Barbara Carminati\IEEEauthorrefmark{1}, Elena Ferrari\IEEEauthorrefmark{1}, John D. Sheehan\IEEEauthorrefmark{2}, Emanuele Ragnoli\IEEEauthorrefmark{2}}
    \IEEEauthorblockA{\IEEEauthorrefmark{1}University Of Insubria, Italy
    \{gsagirlar, barbara.carminati, elena.ferrari\}@uninsubria.it}
    \IEEEauthorblockA{\IEEEauthorrefmark{2}IBM Research - Ireland
    \{john.d.sheehan,eragnoli\}@ie.ibm.com}
}



%


\maketitle
\thispagestyle{plain}
\pagestyle{plain}

\begin{abstract}
From its early days the Internet of Things (IoT) has evolved into a decentralized system of cooperating smart objects with the requirement, among others, of achieving distributed consensus. Yet, current IoT platform solutions are centralized cloud based computing  infrastructures, manifesting a number of significant disadvantages, such as, among others,  high cloud server maintenance costs, weakness for supporting time-critical IoT applications, security and trust issues. Enabling blockchain technology into IoT can help to achieve a proper distributed consensus based IoT system that overcomes those disadvantages. While this is an ideal match, it is still a challenging endeavor. In this paper we take a first step towards that goal by designing Hybrid-IoT, a hybrid blockchain architecture for IoT. In Hybrid-IoT, subgroups of IoT devices form PoW blockchains, referred to as PoW sub-blockchains. Then, the connection among the PoW sub-blockchains employs a BFT inter-connector framework, such as Polkadot or Cosmos. In this paper, we focus on the PoW sub-blockchains formation, guided by a set of guidelines based on a set of dimensions, metrics and bounds. In order to prove the validity of the approach we carry on  a performance and security evaluation.
\end{abstract}



%
\IEEEpeerreviewmaketitle

\input{introduction}

\input{related}
\input{metrics}

\input{integrationEvaluations}

\input{architecture}

\input{performance}
\input{security}

\input{conclusions}

\bibliographystyle{IEEEtran}
\bibliography{biblio}

\end{document}

%% file: introduction.tex
\section{Introduction}\label{sec:introduction}
Internet of Things (IoT) technology is heterogeneously applied to several environments: buildings, automotive, manufacturing, cities, etc., with the potential to make them smarter, more connected, profitable, and efficient. This typically requires the connection, concerted operation and management of a distributed large number of loosely coupled smart devices \cite{ourPaper}, that need to identify and trust each other.
While this should ideally map to a decentralized hardware and software platform, current solutions are mostly based on centralized infrastructures. The disadvantages of that are, among others: high maintenance costs; low interoperability due to restricted data aggregation with other centralized infrastructures; single point of failures (SPOF) against security threats.

Decentralization, if achieved, would have the advantage to reduce the amount of data that are transferred to the cloud for processing and analysis, it would be instrumental to improve security and privacy of the managed data \cite{ourPaper2}, and it would lead to concerted and autonomous operations. For example, in smart home environments \cite{distributedSmartHome}, IoT devices have to autonomously exchange and process data, assure data security, operations accountability, device identification and, last but not least, to collectively and autonomously execute smart homes operations. This, from a distributed systems point of view, means achieving distributed consensus.

A promising decentralized platform for IoT is blockchain. Blockchain is the concept of a distributed ledger maintained by a peer-to-peer network. Its data structure consists of bundled data chunks called blocks, where peers in a blockchain broadcast blocks by exploiting public-key cryptography. Blocks are recorded in the blockchain with exact ordering. Briefly, a block contains: a set of transactions (exchange and transfer of information); a timestamp; a reference to the preceding block that identifies the block's place in the blockchain; an authenticated data structure (e.g., a merkle tree \cite{merkle}) to ensure block integrity.\footnote{Block structure varies in different blockchain protocols, here we list the most common elements.} Modern blockchain protocols, such as Ethereum\footnote{ethereum.org}, possess scripting systems that allow the coding and execution of computing programs on the blockchain itself, referred to as smart contracts \cite{blockchainSurvey}. Different blockchain protocols may employ different methodologies to achieve consensus. For example, some blockchains use \textit{Proof of Work (PoW)} \cite{bitcoinNakamoto}, while others  \textit{Byzantine Fault Tolerant (BFT)} \cite{tendermint}. In PoW blockchains, peers, referred to as block miners, have to use their hardware resources and energy  to solve a cryptographic puzzle as proof of their work, in order to be authorized to generate a new block. Notably, PoW blockchains are able to maintain a relative low throughput while scaling to thousands of nodes in achieving consensus, while BFT blockchains can maintain a relative high throughput with only few nodes. Other consensus protocols are emerging, for example Proof Of Stake (PoS) \cite{blockchainSurvey}, but for the scope of this work we consider mature and well implemented protocols like PoW and BFT.

When it comes to IoT, blockchain can be used to store critical machine-to-machine communications, sent as blockchain transactions, ensuring accountability and security of the stored data. It can also provide identity and proof of provenance of IoT devices with its cryptographic functions. 
While the literature offers some examples of blockchain technology in IoT, such as \cite{blockchainIIoT, managingIoTblockchain, blockchainSecurefirmware}, up to now there is no de facto standard solution.
 
Indeed, one of the biggest challenges in the integration of blockchain into IoT is scalability.
In fact, due to the massive number of devices and resource constraints, deploying blockchain in IoT is particularly challenging. The optimal blockchain architecture has to scale to many IoT devices (they become the peers on the blockchain network), and it should be able to process a high throughput of transactions.

Hybrid-IoT, the platform designed in this work, exploits both PoW blockchains and BFT protocols. 
First, PoW blockchains are used to achieve distributed consensus among many IoT devices, the peers on the blockchain. To measure and qualify that, we define a set of PoW blockchain-IoT integration metrics, and we evaluate the performance of Hybrid-IoT subject to varying blockchain block sizes and block generation intervals, device locations, and number of peers. 
Since we first observed that PoW blockchains containing few hundreds of geographically close IoT devices have high performance (i.e., high transaction throughputs) and low block propagation delays, the first step in Hybrid-IoT consists of generating multiple PoW blockchains. 
Those are generated according to a set of rules, referred to as sweet-spot guidelines, that combine best practices in designing sub-blockchains.
As a second step, Hybrid-IoT leverages on a BFT inter-connector framework ,such as Polkadot\footnote{polkadot.network} and Cosmos,\footnote{cosmos.network} in order to achieve interoperability among sub-blockchains. In this work, we  only deal with the first step, that is, analyzing the performance and security of the PoW sub-blockchain setting.

Furthermore, in  Hybrid-IoT we define three roles for IoT devices according to their capabilities, and test the performance of the system with a set of experiments and simulations. Moreover, we extensively test security of our approach by acknowledging that generating sub-blockchains may generate security vulnerabilities.

We stress that, in this work, we make extensive use of Bitcoin clients and a Bitcoin simulator to conduct the performance analysis.
The Bitcoin client approach is used to test the Hybrid-IoT architecture and design.
The real focus here is the PoW sub-blockchains design with the sweet-spot guidelines and the type of tests and analysis performed here would not differ with other types of PoW protocols that allow smart contracts (this would affect only the types of transactions submitted to the peers).

In the literature, there are few papers targeting application of blockchain to IoT (cfr. Section \ref{sec:related}). 
However, application of blockchain to IoT has been mainly limited to application specific tasks (e.g. firmware updates of IoT devices \cite{blockchainSecurefirmware}), whereas the goal of this paper is to decentralize IoT by exploiting blockchain.

\textbf{Contributions.} Our main contributions are:

$\star$ A set of PoW blockchain-IoT integration metrics to measure performance.

$\star$ A measurement study of the performance of PoW blockchains in IoT.

$\star$ A hybrid blockchain architecture for IoT.

\textbf{Structure.} The remainder of this paper is organized as follows.
Section \ref{sec:related} discusses the related work.
We define PoW blockchain-IoT integration metrics in Section \ref{sec:metrics}.
In Section \ref{sec:integrationEva}, we extensively evaluate PoW blockchain-IoT integration and define sweet-spot guidelines for sub-blockchain generation.
We detail the design of  Hybrid-IoT in Section \ref{sec:architecture}.
In Section \ref{sec:performance}, we evaluate performance of the sub-blockchains.
Security of our approach is discussed in Section \ref{sec:security}, whereas Section \ref{sec:conclusions} concludes the paper.

%% file: related.tex
\section{Related Work}\label{sec:related}
In the literature, there are few instances of application of blockchain to IoT. 
One application of blockchain to IoT is \cite{blockchainIIoT}, where a blockchain platform for industrial IoT (BPIIoT) has been proposed. BPIIoT exploits smart contracts to develop a decentralized manufacturing applications of cloud based manufacturing (CBM). 
Smart contracts have also been exploited in \cite{managingIoTblockchain} in order to manage smart meter data.   Likewise, in \cite{blockchainSecurefirmware} a blockchain based system has been proposed to manage firmware updates of IoT devices. 
\cite{empoweredIoTUsers} exploits blockchain to store access control data, as a data storage system in a multi-tier IoT architecture. 
In \cite{iotchain}, blockchain  and smart contracts are used to secure authorization requests to IoT resources. 
The above mentioned works make use of blockchain  to either execute smart contracts or perform application specific tasks, but not to decentralize IoT systems and achieve autonomous application execution. 

\cite{lsb} proposes a blockchain architecture for IoT containing two layers, namely: smart home layer (centrally managed private ledgers) and a overlay layer (public blockchain).
Resource constrained devices form private ledgers in smart home layer, that are centrally managed by constituent nodes.
Group of constituent nodes select a cluster head operating in the overlay network.
On one hand, this proposal has some similarities with Hybrid-IoT as it includes multiple blockchains.
It relies on distributed trust algorithms to eliminate computational overhead from IoT devices due to PoW solving task. 
However, the proposed architecture does not help with the decentralization of IoT. In fact, IoT devices are centrally managed and connected to one constituent node that does not take part in the distributed consensus.

Differently from previous works, the Tangle\footnote{\url{iota.org/IOTA_Whitepaper.pdf}} protocol implements a global distributed ledger for IoT by using Directed Acyclic Graph  generated by transactions as a blockchainless approach.  Tangle is designed as a cryptocurrency for IoT to make micro-payments possible, it does not provide an architecture or data structure to decentralize IoT and it is not Turing complete to allow scripting and smart contracts.

%% file: metrics.tex
\section{PoW Blockchain-IoT Integration Metrics}\label{sec:metrics}
We identify five relevant dimensions that an optimal blockchain PoW implementation for IoT should be subject to: \textit{scalability}, \textit{security}, \textit{decentralization}, \textit{efficiency} as observed metrics, and \textit{network bandwidth} as a controlled parameter. 
In what follows, we analyze those dimensions  (see Table \ref{tab:metrics} for a summary).

\textbf{Scalability.} 
Scalability in IoT is the capacity to be changed in size or scale in terms of number of devices,  hardware characteristics and functional and non-functional requirements, while maintaining quality of performance. For blockchain, this translates to have a peer to peer network that can scale up in terms of number of peers and throughput, as number of transactions per unit of time.

\textbf{Security.} 
Security is a critical dimension in IoT, especially considering recent large scale attacks ,like Mirai and WannaCry.\footnote{siliconrepublic.com/machines/iot-devices-botnets-autonomous-cars}  While in this work we do not deal with device intrusions, the issue of data integrity for IoT devices is an important problem to be solved \cite{blockchainDatIntegrityIoT}. While data integrity is by design preserved by a PoW blockchain, the issue of the longer chain attack still exists \cite{onTheSecurityPoW}. In order to measure this, we consider the maximum amount of total work in the PoW sub-blockchain as a metric.

\textbf{Decentralization.} 
Decentralization in IoT is critical to improve security and privacy and achieve autonomous execution, as noted in Section \ref{sec:introduction}.  In peer-to-peer overlay networks, like blockchains, decentralization is measured by the number of properly functioning peers \cite{scalingDecentralized}. In a blockchain, a peer needs to be up to date with the most recent block before generating a new block to be accepted by blockchain consensus. Hence, we define the metrics for measuring decentralization as the number of functioning peers on the network. We also define a lower bound of functioning peers, to be 90\% of the total, to guarantee proper functionality of the blockchain for its IoT application.

\textbf{Efficiency.} 
Efficiency in IoT can be defined  as an optimal utilization of hardware resources and energy. 
Therefore, in order to achieve that, the IoT devices on the blockchain should optimally utilize resources and energy to maintain and progress the blockchain. Among others, an obstacle to that is the issue of forks and stale blocks in PoW blockchains \cite{onTheSecurityPoW}. Specifically, stale  blocks do not contribute to the security of the blockchain and transactions in stale blocks are considered as unprocessed by the network, requiring wasted effort to generate them.\footnote{In Ethereum blockchain they are included to the blockchain as uncle blocks, however they do not count towards total difficulty of the blockchain \cite{onTheSecurityPoW}.} Hence, we define our metrics for efficiency as the stale block generation ratio and we establish a upper bound for performance to be $\approx$1\%.

\textbf{Network bandwidth.} 
Network bandwidth is a one to one map between the IoT network and its corresponding blockchain network. It is defined by the the IoT devices downlink and uplink rates. For example IEEE 802.15.4 and NarrowBand-IoT standards set 250 Kbps data transfer peak rates for machine to machine communication, whereas in LTE Cat M1 and LTE Cat 0 standards it is 1 Mbps. In this work, in order to avoid network overloads and consequent bottlenecks with high information traffic, we set an upper bound of 250 Kbps as total of uplink and downlink rates.

\begin{table}  
\centering
\setlength{\tabcolsep}{10pt}
\begin{tabular} {cc} 
\toprule
\textbf{Dimensions} & \textbf{Metrics} \\
\cmidrule(lr){1-1} \cmidrule(lr){2-2}  
{\textit{Scalability}} 	   		     		         & $\diamond$ Maximum no of IoT devices as peers \\
\textbf{}		   			  		    & $\diamond$ Maximum transaction throughput\\
\cmidrule(lr){1-1} \cmidrule(lr){2-2} 
{\textit{Security}} 	            & $\diamond$ Maximum work in the blockchain  \\
\cmidrule(lr){1-1} \cmidrule(lr){2-2} 
{\textit{Decentralization}}    &$\diamond$ $
\displaystyle\frac{\mbox{90\% block propagation time}}{\mbox{block generation interval}}$ $\leq$ 1 \\
\cmidrule(lr){1-1} \cmidrule(lr){2-2} 
{\textit{Efficiency }}  & $\diamond$ Stale block generation rate  $\approx$ 1\%\\
\cmidrule(lr){1-1} \cmidrule(lr){2-2} 
{\textit{Network bandwidth}} 		  & $\diamond$ Avg network traffic of a device $\leq$ 250 Kbps \\
\bottomrule             
\end{tabular}
\caption{PoW Blockchain - IoT Integration Metrics} \label{tab:metrics} 
\end{table}

%% file: integrationEvaluations.tex
\section{PoW Blockchain-IoT Integration Evaluations}\label{sec:integrationEva}
In this section, we evaluate the performance of the integration of PoW blockchains in IoT, subject to the dimensions and metrics defined in Section \ref{sec:metrics}.
To this end, we use and further extend (by adding different device location setups) the Bitcoin simulator\footnote{github.com/arthurgervais/Bitcoin-Simulator} presented in \cite{onTheSecurityPoW}, (see Section \ref{sec:simulatorSettings}). We perform three evaluations (see Section \ref{sec:evalutionResults}): one by varying block size and block generation intervals (see Section \ref{sec:eva1}); one by varying device location (see Section \ref{sec:eva2}); one by varying the number of IoT devices (see Section \ref{sec:eva3}). We present results as an average of 5 experimental runs.
We use the findings of this section to define the concept of sweet-spot guidelines that drives the generation of sub-blockchains (see Section \ref{sec:sweetspot}).

\subsection{Simulator Setting}\label{sec:simulatorSettings}
The Bitcoin simulator is built on \textit{ns-3} discrete-event network simulator.
It allows to model a Bitcoin network with a set of consensus and network parameters such as: block generation interval; block size; number of nodes.
Connections between nodes are established using point-to-point channels, by considering latency and bandwidth as the two main characteristics (cfr.  \cite{onTheSecurityPoW} for further information).
We have extended the simulator with three different device location setups, namely \textit{the Netherlands}, \textit{Europe}, and \textit{World}, by adopting real world network latency data.\footnote{wondernetwork.com/pings}
In the Netherlands setup, devices are located in  six cities of the Netherlands: \textit{Alblasserdam, Amsterdam, Dronten, Eindhoven, Rotterdam} and \textit{The Hague}. 
In the Europe setup, devices are located in six European cities: \textit{Brussels, Athens, Barcelona, Izmir, Lisbon} and \textit{Milan}.
Finally, in the World setup devices are located in 7 globally distributed cities: \textit{Dhaka, Hangzhou, Istanbul, Lagos, Melbourne} and \textit{San Diego}. We equally distribute regular and miner among the cities in the respective setups. 

In order to use the simulator for our evaluations we categorize IoT devices within two roles: miners and regular devices. The number of connections per miner device and regular device follows the distribution as in \cite{discoveringTopology}.
Regular devices only check and propagate the blocks they receive, whereas miner devices also generate new blocks. The ratio of miner over number of nodes is set to ca 7\%, with the remainder taking the role of regular devices. This is justified by some Bitcoin statistics \cite{discoveringTopology} and by the fact that we consider only a small subset of IoT devices to have enough resources to take part in the mining process. 

Network latency plays a critical role in performance due to the intrinsic nature of peer to peer information propagation (i.e., block and transaction). 
Hence, to evaluate how geographical locations of the devices affect network latency, we exploit \textit{the Netherlands}, \textit{Europe}, and \textit{World}  device location settings of the simulator.

Bandwidth capacities of IoT devices obviously affect information propagation time in the blockchain.
To have a realistic bandwidth setup, we adopt the bandwidth benchmarks of Raspberry Pi devices.\footnote{pidramble.com/wiki/benchmarks/networking}
Hence, we adopt an upper bandwidth limit of 100 Mbps (variations within that limit are allowed due to  connection type) and we realistically simulate bandwidth capacities with a distribution from testmy.net.\footnote{testmy.net/country} That results in a varying download bandwidth between 0.1 Mbps and 100 Mbps with a 5 Mbps average, and a varying upload bandwidth between 0.02 to 20 Mbps with a 1Mbps average. 

\subsection{Evaluation Results}\label{sec:evalutionResults}
\subsubsection{Evaluation \rom{1}: Block sizes and block generation intervals}\label{sec:eva1}
We evaluate the effect of block sizes and block generation intervals with the simulator with the Netherlands setup. We adopt a six block generation cycle with the following intervals: 10 minutes, 5 minutes, 1 minute, 30 seconds, 10 seconds, and 5 seconds.
For every block generation cycle, we vary block sizes as:  10 KB, 50 KB, 100 KB, 500 KB, 1 MB, 5 MB, 10 MB. We fix the number of IoT devices to 250, with 18 devices with miner roles, according to the 7\% ratio in Section \ref{sec:simulatorSettings}.
Experiment results are presented in Table \ref{tab:eva1}.

\textbf{Network bandwidth.} Not surprisingly, using bigger blocks and/or having short block generation intervals increase the average network traffic.
In that, big blocks (e.g., 5 MB) comply with the network bandwidth metric's bound when block generation interval is long enough (e.g., 5m), whereas for small blocks (e.g., 10 KB) even short block generation intervals (e.g., 5s) are suitable.

\textbf{Security.} Obviously, using shorter block generation intervals increases the number of blocks generated.
However, we observe that this is not proportional, especially when the block size is bigger than 100 KB. Similarly, in experiments with 1 minute or shorter block generation interval settings, increasing the block size decreases the number of generated blocks.
This is due to bandwidth exhaustion of devices.
Therefore, according to the bounds of security metric, using small blocks (e.g., 10 KB) in short block generation intervals (e.g., 5s) is more appropriate to increase number of genuine blocks.

\textbf{Decentralization.} According to the decentralization metric's bounds, 90\% block propagation time should be lower than block generation interval. 
Due to their restricted bandwidth capabilities, IoT devices have to spend more time to propagate big blocks, and that in turn breaches the 90\% block propagation time bound. 
In parallel, we observe that,  when using big blocks (e.g., 10 MB), block generation interval should be long enough (e.g., 10m) to satisfy the decentralization lower bound. 
For example, when small blocks (e.g. 10 KB) are used, the decentralization bound can be satisfied with shorter block generation intervals (e.g., 5s). 
Therefore, in order to achieve decentralization, block sizes and block generation intervals should be set carefully. 

\textbf{Efficiency, scalability.} Short block generation intervals and/or using big blocks leads to  higher stale block rates, as bandwidth resources of IoT devices are exhausted in propagating the blocks. In order to achieve low stale block rates, with a  short block generation interval setup, only small blocks can be used. Bigger blocks (e.g., 1 MB) can be used with long block generation intervals. 
The bigger the block is, the longer block generation should be used to satisfy the low stale block generation bound. 
Moreover, we observe that block sizes bigger than 1 MB are not suitable for IoT, since it leads to high stale block rates, even with a long block generation interval setup.
Achieving a low stale block rates positively impacts transaction throughput. In our experiments the highest throughput achieved is 30.1 transaction per second in using 500 KB blocks with 1 minute block generation interval setting with 1.71\% stale block rate.

\textbf{Findings:} blocks smaller than 1 MB should be used; block generation intervals should be as short as possible; block size and block generation intervals should be set carefully to ensure low stale block rates and high decentralization. 


\begin{table}
    \centering
    \scriptsize
    \begin{tabular} {*{9}{c}} 
        \hline
         \textbf{Block}  & {\bf Block.} & {\bf Total }    &  \textbf{Stale}   & \textbf{Genuine}  &{\bf Stale}    &{\bf 90\%}       &{\bf Avg }   & {\bf Thrghpt} \\
		 \textbf{Size}   & {\bf Gen.}   &\textbf{Blocks}  & \textbf{Blocks}   & \textbf{Blocks}   &\textbf{Rate}  &\textbf{Prop.}   &\textbf{Traffic} & {\bf (TX/s)}\\ 
         \textbf{}       & {\bf Intrvl(s)}  & 			  &	 			      &                   &               &\textbf{Delay(s)}&\textbf{(Kbps)} & \\
     	  \hline
                    	 & {\bf  10m}& { 10.8} 		    & {0.43} 	     &{ 10.4}  			  &  { 3\%}	     & { 360}  &{ 276}    & { 69.3}  \\
                    	 & {\bf  5m}& { 18.8} 			& {0.9} 		 &{ 17.9}  			  &  { 8.83\%}	 & { 755}  &{ 723}    & { 119.5}  \\
         \textbf{10 MB}  & {\bf  1m}& { 45.6} 	   		& {16} 		     &{ 26.93}  		  &  { 35.07\%}	 & { 2162}  &{ 21215}  & { 197.5}  \\
                    	 & {\bf  30s}& { 51.2} 	        & {26.6} 		 &{ 24.6}  			  &  { 47.99\%}	 & { 2412} &{ 49520}   & { 164.1}  \\
                    	 & {\bf  10s}& { 57.7} 		    & {41.2} 		 &{ 16.5}  			  &  { 71.38\%}	 & { 2560} &{ 151046}  & { 110.2}  \\
                    	 & {\bf  5s}& { 64.2} 			& {48.2} 		 &{ 16}  			  &  { 75.00\%}	 & { 2665} &{ 273777}  & { 107.1}  \\                  
        \hline
                 		 & {\bf  10m}& { 10.2}	        & {0.26}		 &{ 9.9}  			  &  { 2.6\%} 		 & { 168}   &{ 134}   & { 33.1}  \\
                 		 & {\bf  5m}& { 19.9}		    & {1}		     &{ 18.9}  			  &  { 5.3\%} 	 	 & { 180}   &{ 288}   & { 63}  \\
         \textbf{5 MB}	 & {\bf  1m}& { 67.3}		    & {12.1}		 &{ 55.2}  			  &  { 17.99\%} 	 & { 1888}  &{ 8718}  & { 184}  \\
                 		 & {\bf  30s}& { 73.4}		    & {26.8}		 &{ 46.6}  			  &  { 36.52\%} 	 & { 2105}  &{ 28528} & { 155.5}  \\
                 		 & {\bf  10s}& { 84}		    & {56.5}		 &{ 27.5}  			  &  { 67.18\%} 	 & { 2472} &{ 100255}& { 92}  \\
                 		 & {\bf  5s}& { 91.4}		    & {68.5}		 &{ 22.9}  			  &  { 74.91\%} 	 & { 2512} &{ 190671}& { 76.4}  \\              
        \hline
         				 & {\bf  10m}& { 12.3} 			& {0} 		 	 &{ 12.3}     		  &  { 0\%} 	 & { 31}   &{ 26} & { 8.2} \\
         				 & {\bf  5m}& { 22.3} 		    & {0} 		     &{ 22.3}     		  &  { 0\%} 	 & { 32}   &{ 53}   & { 14.9} \\
         \textbf{1 MB}   & {\bf  1m} & { 92.4} 			& {3.4} 		 &{ 89}     		  &  { 3.71\%} 	 & { 37}   &{ 438} & { 59.3} \\
         			 	 & {\bf  30s}& { 165.9}			& {8.5} 		 &{ 157.4}     	 	  &  { 5.15\%} 	 & { 818}   &{ 3243}  & { 104.9} \\
         			 	 & {\bf  10s}& { 219.6}         & {94.1} 		 &{ 125.5}     	  	  &  { 42.86\%}  & { 1812}  &{ 30259}   & { 83.7} \\
         			 	 & {\bf  5s} & { 232.5} 		& {128.7} 		 &{ 103.8}     	   	  &  { 55.37\%}  & { 2183} &{ 69059}  & { 69.2} \\ 
        \hline
         	 		 	 & {\bf  10m}& { 9.6}          & {0} 	       	 &{ 9.6}              &  { 0\%}    	  & { 15}    &{ 13}  & { 3.2}\\
         	 		   	 & {\bf  5m}& { 18.6}          & {0} 			 &{ 18.6}     		  &  { 0\%}    	  & { 15}    &{ 26}   & { 6.2}\\
         \textbf{500 KB} & {\bf  1m}& { 92.1}          & {1.6} 	     	 &{ 90.5}             &  { 1.71\%}    & { 17}    &{ 136}  & { 30.1}\\
         	 		     & {\bf  30s}& {165.2}         & {9.2} 	     	 &{ 156}              &  { 5.56\%}    & { 18}   &{ 639} & { 52}\\
         	 		     & {\bf  10s}& {346.5}         & {101.4} 	 	 &{ 245.1}            &  { 29.25\%}   & { 1665}  &{ 14762}  & { 81.7}\\
         	 		     & {\bf  5s}& { 350.6}         & {161.1} 	 	 &{ 189.5}            &  { 45.96\%}   & { 1972} &{ 41378}   & { 63.2}\\
         
        \hline
         	     		& {\bf  10m}& { 9.3}           &{0} 	     	 &{ 9.3}               &  { 0\%}       & { 3.2}     &{ 2}    & { 0.6} \\
                 		& {\bf  5m}& { 23}             &{0} 		 	 &{ 23}     		   &  { 0\%}       & { 3.2}     &{ 5}    & { 1.5} \\
         \textbf{100 KB}& {\bf  1m} & { 99}            &{0} 	     	 &{ 99}                &  { 0\%}       & { 3.2}     &{ 27}   & { 6.6} \\
         	     		& {\bf  30s}& { 186.4}         &{4.4} 	         &{ 182}               &  { 2.35\%}    & { 3.2}     &{ 54}   & { 12.1} \\
         	     		& {\bf  10s}& { 537.3}         &{22.8} 	         &{ 514.5}             &  { 4.25\%}    & { 3.4}     &{ 447}  &{ 34.3} \\
         	     		& {\bf  5s}& { 954.5}          &{124.5} 	     &{ 830}               &  { 13.04\%}   & { 99}    &{ 7249} & { 55.3} \\
        \hline 
         	 	 		& {\bf  10m}& { 11}            &{0}         	 &{ 11}               &  { 0\%}       & { 1.6}    &{ 1}   & { 0.4} \\
         	 	 		& {\bf  5m}& { 18.6}           &{0} 		 	 &{ 18.6}     		  &  { 0\%}   	  & { 1.6}    &{ 2}   & { 0.6} \\
         \textbf{50 KB} & {\bf  1m}& { 96.3}           &{0} 	     	 &{ 96.3}             &  { 0\%}       & { 1.6}    &{ 14}  & { 3.2} \\
         	 	 		& {\bf  30s}& { 187.0}         &{0.7}           &{ 186.3}            &  { 0.35\%}    & { 1.6}    &{ 28}  & { 6.2} \\
         	 	 		& {\bf  10s}& { 562.0}         &{10.2}           &{ 551.8}            &  { 1.82\%}    & { 1.7}    &{ 84}  & { 18.4} \\
         	 	 		& {\bf  5s}& { 1120.4}         &{43.4}           &{ 1077}           &  { 3.87\%}    & { 1.8}      &{ 931} & { 35.9} \\      
        \hline      
         	      	   & {\bf  10m}& { 10.3}         	 &{0}           &{ 10.3}             &  { 0\%}    & { 0.4}    &{ 0.3}  & { 0.1}  \\
         	      	   & {\bf  5m}& { 21.6}             &{0} 		    &{ 21.6}     		 &  { 0\%}    & { 0.4}    &{ 0.7}  & { 0.2}  \\
         \textbf{10 KB}& {\bf  1m}& { 101}              &{0} 	        &{ 101}              &  { 0\%}    & { 0.4}    &{ 3.5}  & { 0.7}  \\
         	      	   & {\bf  30s}& { 193.3}           &{0}           &{ 193.3}            &  { 0\%}    & { 0.4}    &{ 7}    & { 1.3}  \\
         	      	   & {\bf  10s}& { 598.6}           &{0}           &{ 598.6}            &  { 0\%}    & { 0.4}    &{ 21}   & { 4}  \\
         	      	   & {\bf  5s}& { 1166.3}           &{19.9}         &{ 1146.4}             &  { 1.71\%} & { 0.4}    &{ 42}   & { 7.6}  \\      
        \hline    
    \end{tabular}
\caption{Evaluation \rom{1}: Block sizes and block generation intervals} \label{tab:eva1}
\end{table}

\subsubsection{Evaluation \rom{2}: Device Locations}\label{sec:eva2}
We evaluate the effect of device locations by varying the network latency among IoT devices.
In order to simulate that, we use the Bitcoin simulator with three location settings (the Netherlands, Europe, and World), as in Section \ref{sec:simulatorSettings}. Since from Evaluation \rom{1}, the optimal block size should be less than or equal than 1 MB, the block size is fixed at 500 KB on average. We adopt a six block generation cycle with the following intervals: 10 minutes, 5 minutes, 1 minute, 30 seconds, 10 seconds, and 5 seconds. We fix the number of IoT devices to 250, where 18 of them are miners. Experiment results are presented in Table \ref{tab:eva2}. 

\textbf{Network bandwidth, security.} For all location setups, in each block generation interval setting, average network traffic per device and number of generated genuine blocks are highly correlated.
Particularly, only 1 minute or longer block generation intervals comply with the bound for the the network bandwidth metric (the average network traffic should be less than 250 Kbps) for all location settings.
Hence, a 1 minute block generation interval is the most suitable according to the security metric bound, since it has the highest number of genuine blocks. With those, every locations setup shows a similar behavior.

\textbf{Scalability, decentralization, efficiency.}
For any setup that we tried the outcome with shortest block propagation delays, lowest stale block rates, and highest transaction throughputs is the Netherlands setup. For example, with 1 minute block generation interval, a PoW blockchain using the Netherlands setup achieves a throughput of 30.1 per second and complies with the efficiency bound (stale block rate is 1.71\%) and decentralization bound (90\% block propagation time is 17 seconds). Whereas, in Europe and World settings, the block generation interval needs to be at least 5 minutes to satisfy the same bounds. With those, both the Europe and World setups can only achieve a throughput of 5 transaction per second. 

\textbf{Findings:} blockchains containing IoT devices that are geographically close to each other achieve higher throughput with low stale block rates.

\begin{table} 
    \centering
    \scriptsize
    \begin{tabular}{*9{c}}
        \hline
         \textbf{Block  }        & {\bf  Sce.}& {\bf Total }  		&  \textbf{Stale} 		& \textbf{Genuine} 		 &{\bf Stale}   	&{\bf Mean}       &{\bf Avg }   & {\bf Thrghpt} \\
		 \textbf{Gen.}           & {\bf  }   &\textbf{Blocks}		& \textbf{Blocks} 		& \textbf{Blocks}   	&\textbf{Rate} 		&\textbf{Delay}   &\textbf{Traffic} & {\bf (TX/s)}\\ 
         \textbf{Intrvl(s)}      & {\bf  }   & 			  			&	 			    	&                   	&              		&\textbf{(s)}&\textbf{(Kbps)} & \\
     	  \hline
          						 & {\bf  N}	&  9.6              &{0} 	       & 9.6			   &  { 0\%}	     & { 7}       & { 13}    &{3.2}  \\
          \textbf{10m}  	     & {\bf  E}	&  9.9			    &{0} 		   &{ 9.9}  		   &  { 0\%}	     & { 16}      & { 13}    &{3.3}\\
                               	 & {\bf  W}	&  9.9 			    &{0} 		   &{ 9.9}  		   &  { 0\%}	     & { 20}      & { 13}	 &{3.3}\\
          \hline
         			     		 & {\bf  N}	& { 18.6}          & {0} 		   &{ 18.6}  		   &  { 0\%} 	     & { 6}       & { 26}     &{ 6.2}  \\
         \textbf{5m}    	     & {\bf  E}	& { 16.4}		   & {0}		   &{ 16.5}  		   &  { 0\%} 		 & { 13}      & { 27}     &{ 5.5}\\
        			    		 & {\bf  W}	& { 15.5}		   & {0}		   &{ 15.5}  		   &  { 0\%} 	     & { 17}      & { 27}     &{ 5.2}\\
		 \hline       
         						 & {\bf  N}& { 92.1}           & {1.6} 	     &{ 90.5}     		   &  { 1.71\%} 	 & { 17}    & { 136}     &{30.1}\\
         \textbf{1m}		     & {\bf  E}& { 93.6} 		   & {4.8} 		 &{ 88.8}     		   &  { 5.14\%} 	 & { 18}    & { 140}     &{29.6}\\
                                 & {\bf  W}& { 96.5} 		   & {7.9} 		 &{ 88.6}     		   &  { 8.22\%} 	 & { 20}    & { 140}     &{29.5}\\
         \hline
                  			 	 & {\bf  N}& { 165.2}         & {9.2} 	     &{ 156}              &  { 5.56\%}    & { 18}      & { 639}      & { 52}\\                              
         \textbf{30s} 	 		 & {\bf  E}& { 170.5}         & {21.9} 	     &{ 148.6}            &  { 12.87\%}   & { 38}      & { 527}      &{ 49.5}\\
                            	 & {\bf  W}& { 169.5}         & {23.9} 	     &{ 145.6}            &  { 14.11\%}   & { 52}      & { 592}      &{ 48.5}\\
        \hline
         					     & {\bf  N}& { 346.6}         &{101.4} 	     &{ 245.2}            &  { 29.25\%}   & { 314}    & { 14762}  & { 81.7} \\
         \textbf{10s}    		 & {\bf  E}& { 314}           &{104.6} 	     &{ 209.4}            & { 33.31\%}    & { 355}    & { 15237}  &{ 69.8}\\
                                 & {\bf  W}& { 331}           &{136.8} 	     &{ 194.2}            &  { 41.34\%}   & { 392}    & { 16777}  &{ 64.7}\\
         \hline 
         					 	 & {\bf  N}& { 350.7}         &{161.2} 	     &{ 189.5}            &  { 45.96\%}   & { 815}    & { 41378} &{ 63.2} \\
         \textbf{5s} 	 		 & {\bf  E}& { 301.2}         &{156}         &{ 145.2}            &  { 51.80\%}   & { 918}    & { 43931} &{ 48.4}\\
                             	 & {\bf  W}& { 303.3}         &{161.1}       &{ 142.2}            &  { 53.12\%}   & { 1000}   & { 45021} &{ 47.4}\\
        \hline       
    \end{tabular}
\caption{Evaluation \rom{2}: Device Locations} \label{tab:eva2}
\end{table}

\subsubsection{Evaluation \rom{3}: Number of IoT devices}\label{sec:eva3}
We evaluate the effect of varying the number of IoT devices with two experiment types: experiment (A): we fix the PoW difficulty; experiment (B): we fix block generation interval. Each setup is run for 100 minutes.
In both types we vary the number of IoT devices from 83 to 1250 and assumed a fixed block size of 500 KB. 
In PoW blockchains, the block generation interval depends on the ratio of the difficulty of the PoW puzzle over the total mining power of the system \cite{bitcoinNakamoto}.  Hence, with the PoW difficulty fixed, we vary the block generation intervals inversely proportionally to the number of miners. On the other hand, with a fixed block generation interval of 1 minute, the difficulty of the PoW puzzle is varied proportionally to number of miners (the difficulty of the PoW puzzle is $\alpha$ for 6 miners and 15$\alpha$ for 90 miners). 

\textbf{Experiment (A).} Results are in Table \ref{tab:case3.1}.

\textbf{Metrics.} 
Having more IoT devices with shorter block generation intervals leads to generate more blocks, leading to an increase in throughput and average network traffic per device. That causes extensive bandwidth consumption that generates long block propagation delays, which leads to high stale block rates. Hence, experimental variations containing 83, 166 and 250 devices satisfy the efficiency, network bandwidth, and decentralization bounds. 
When it comes to scalability and security bounds, scenario containing 250 devices is the optimal setups as it produces more genuine blocks, and achieves the highest throughput and scales to more devices.

\begin{table}
    \centering
    \scriptsize
    \begin{tabular}{*9{c}}
        \hline
         \textbf{No of}         & {\bf Block}  & {\bf Total }  	&  \textbf{Stale} 		& \textbf{Genuine} 		 &{\bf Stale}   	    &{\bf 90\%}       &{\bf Avg }       & {\bf Thrghpt} \\
		 \textbf{Miners/}      	& {\bf Gen.}   &\textbf{Blocks}	& \textbf{Blocks}       & \textbf{Blocks}   	 &\textbf{Rate} 		&\textbf{Delay}   &\textbf{Traffic} & {\bf (TX/s)}\\ 
         \textbf{Total}         & {\bf Intrvl(s)} & 			&	 			    	&                    	 &              		&\textbf{(s)}     &\textbf{(Kbps)}  & \\
     	  \hline
         \textbf{6/83}     		& {\bf 3m}		& {29.9}      			&{0}	  	    		&{29.9}   		 	 	&  { 0\%}	    	&{7}           	  &{ 44}       & { 9.9}\\
         \hline
         \textbf{12/166}    	& {\bf 1.5m}	& {76.7}      			&{1.3}	    			&{75.4}	 		 	 	&  { 1.9\%}	 	    &{8}          	  &{ 88}       & { 25.1}\\
         \hline    
         \textbf{18/250}    	& {\bf  1m}      & { 92.1}              & {1.6} 	     	    &{90.5}                 &  { 1.71\%}        &{17}             &{ 136}      & { 30.1}\\            \hline 
         \textbf{36/500}    	& {\bf 30s}	    & {177.7}     			&{15.6}	    			&{162.1}			 	&  { 8.8\%}	      	&{41}     		  &{ 1168}		& { 54}\\
         \hline 
         \textbf{54/750}		& {\bf 20s}	    & {237.2}     			&{32.9}	    			&{204.3}	 		 	&  { 13.87\%}	  	&{147}     		  &{ 3485}		& { 68.1}\\
         \hline 
         \textbf{ 72/1000} 		& {\bf 15s}	    &{216.5}      			&{39.4}	    			&{177.1}			 	&  { 18.2\%}	  	&{291}     		  &{ 5791}		& { 59}\\
         \hline 
         \textbf{90/1250}   	& {\bf 12s}	    &{212}        			&{47.5}	    			&{164.5}	 		 	&  { 22.43\%}	  	&{498}     		  &{ 8265}		& { 54.8}\\
         \hline       
    \end{tabular}
\caption{Evaluation \rom{3}: Number of IoT Devices - Experiment (A): Fixed difficulty setting} \label{tab:case3.1}
\end{table}

\textbf{Experiment (B).} Results are in Table \ref{tab:case3.2}.

\textbf{Metrics.} In all the experimental variations, 90\% block propagation times are less than 1 minute block generation interval, thus satisfying the decentralization bound. Similarly,  average network traffic per device is less than 250 Kbps, satisfying the network bandwidth bounds for all experimental variations. However, only experimental variations containing 83, 166 and 250 devices satisfy the efficiency bound with low stale block rates. Among them, experiment containing 250 devices is the optimal setup according to the security and scalability bounds, as it achieves the highest throughput and scales to more devices.

\begin{table}
    \centering
    \scriptsize
    \begin{tabular}{*9{c}}
        \hline
         \textbf{No of}          & {\bf Total }  	& \textbf{PoW}&  \textbf{Stale} 		& \textbf{Genuine} 		 &{\bf Stale}   	    &{\bf 90\%}       &{\bf Avg }       & {\bf Thrghpt} \\
		 \textbf{Miners/}        &\textbf{Blocks}	&\textbf{Puzzle}& \textbf{Blocks}       & \textbf{Blocks}   	 &\textbf{Rate} 		&\textbf{Delay}   &\textbf{Traffic} & {\bf (TX/s)}\\ 
         \textbf{Total}          & 					&\textbf{Difficulty}&	 			    	&                    	 &              		&\textbf{(s)}     &\textbf{(Kbps)}  & \\
     	  \hline
         \textbf{6/83}           & {96.1}    & $\alpha$ &{0,8}         &{95.3}	  &  { 0.85\%}	     &{13}        &{ 135.76}       & { 31.7}\\
         \hline
         \textbf{12/166}         & {96.3}    & 2$\alpha$ &{1.8}	       &{94.5}	  &  { 1.19\%}	     &{14}        &{ 133.37}       & { 31.1}\\
         \hline         
         \textbf{18/250}         &{ 92.1}     & 3$\alpha$ &{1.6} 	   &{ 90.5}   &  { 1.71\%}       &{17}        &{ 136}          & { 30.1}\\           
         \hline 
         \textbf{36/500}         & {93.03}    & 6$\alpha$ &{3.99}	   &{89.04}	  &  { 4.29\%}	     &{26}        &{ 122.99}       & { 32.86}\\
         \hline  
         \textbf{54/750}         & {93.41}    & 9$\alpha$ &{4.49}	   &{88.92}	  &  { 4.8\%}	     &{28}        &{ 102.03}       & { 29.64}\\
         \hline 
         \textbf{72/1000}        & {93.03}    & 12$\alpha$ &{4.42}	   &{88.61}	  &  { 4.75\%}	     &{28}        &{ 102.99}       & { 29.53}\\
         \hline 
         \textbf{90/1250}        & {92.77}    & 15$\alpha$ &{4.98}	   &{87.78}	  &  { 5.36\%}	     &{39}        &{ 107.01}       & { 29.26}\\
         \hline       
    \end{tabular}
\caption{Evaluation \rom{3}: Number of IoT Devices - Experiment (B): Fixed interval setting} \label{tab:case3.2}
\end{table}

\textbf{Findings:} PoW blockchains containing few hundreds of IoT devices achieve higher transaction throughput; the optimal number of IoT devices as blockchain peers is around 250.

\subsection{Sweet-spot Guidelines}\label{sec:sweetspot}
After Evaluations \rom{1},\rom{2} and \rom{3}, we can conclude that PoW blockchains containing few hundreds of IoT devices in close geographical proximity achieve the highest performance. Therefore, in order to design a blockchain architecture for IoT, we propose to deploy multiple PoW blockchains as sub-blockchains for IoT, organized according to pools of IoT devices. We adopt the following guidelines, referred to as sweet-spot:

$\bullet$ Sub-blockchains should contain few hundreds of IoT devices.

$\bullet$ Sub-blockchains should contain IoT devices that are geographically close and frequently communicating with each other. 

$\bullet$ Block size and block generation intervals should be set to ensure low stale block rates, and high decentralization and scattering of mining power.

$\bullet$ Blocks smaller than or equal to 1 MB should be used.

$\bullet$ Block generation interval should be as short as possible.

In the next section, we design the architecture of Hybrid-IoT, based on  the sweet spot guidelines, by leveraging on multiple PoW sub-blockchains.

%% file: architecture.tex
\section{Hybrid-IoT: Hybrid Blockchain Architecture for IoT}\label{sec:architecture}
Hybrid-IoT consists of multiple PoW sub-blockchains that achieve distributed consensus among IoT devices that are peers on the blockchain. 
Sub-blockchains are generated according to the sweet-spot guidelines defined in Section \ref{sec:sweetspot}. 
In order to connect the sub-blockchains, Hybrid-IoT uses a BFT inter-connector framework (e.g.,  Polkadot and Cosmos) that guarantees inter-blockchain transactions.

\begin{figure*} [h] 
	\centering
	\includegraphics[width=\textwidth]{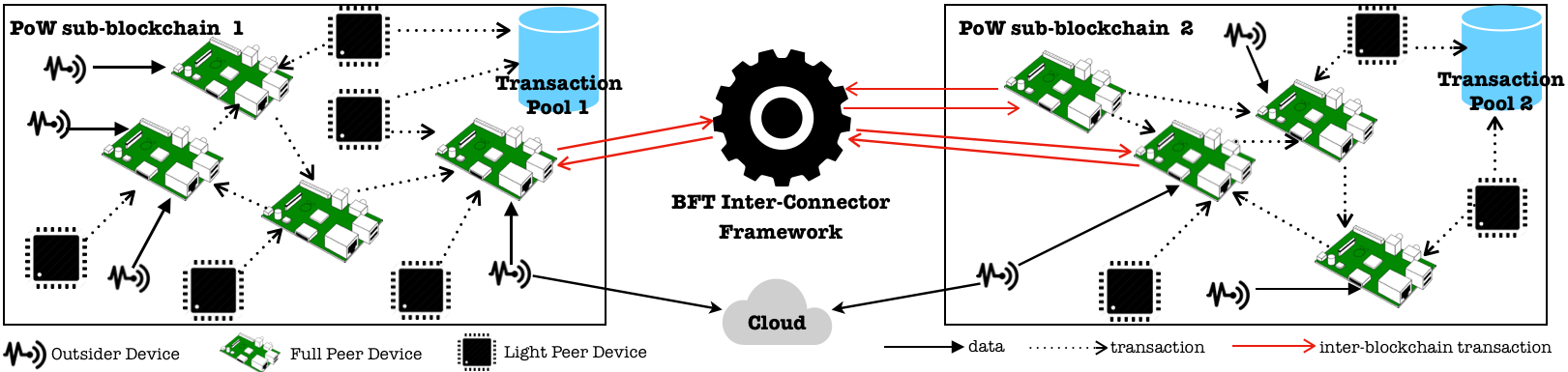}
	\caption{Hybrid-IoT} \label{fig:architecture}
\end{figure*}

\textbf{System execution.}
The transaction flow in Hybrid-IoT is as follows: transactions on the PoW sub-blockchains are processed and included in blocks that are added to their respective sub-blockchain upon PoW consensus;  when a transaction among two distinct sub-blockchains happens, that is picked by the BFT inter-connector framework; the BFT inter-connector framework checks the transaction correctness and authenticity; after a positive response, the BFT inter-connector framework transfers the transaction to the target sub-blockchain's transaction pools that hold unprocessed transactions; last, the transaction is processed and included in a newly generated block in the respective sub-blockchain, upon PoW consensus.
The reasons for the choice of a BFT inter-connector framework lies in the intrinsic capability of BFT consensus protocols to achieve high throughput with a low number of peers.\footnote{For example, Tendermint protocol used by Cosmos network is able to process thousands of transactions per second \cite{tendermint}, whereas PoW sub-blockchains are able to process few dozens of transactions according to our evaluations presented in Section \ref{sec:integrationEva}.} Hence, that should allow to connect few sub-blockchains with an adequate throughput for inter-blockchain transactions. Moreover, by maintaining low latency in the transmission of inter-blockchain transactions, the BFT inter-connector framework allows the connection of a new sub-blockchain without deferring application execution. 
An example of Hybrid-IoT architecture containing two sub-blockchains is shown in  Figure \ref{fig:architecture}.

\textbf{Consensus participation.}
Blockchains can be categorized into two groups subject to the type of peer access control: \textit{permissioned} and \textit{permissionless} blockchains \cite{blockchainOverview}.
In permissionless blockchains, all peers can take part in the consensus, whereas in permissioned blockchains, only pre-defined peers can take part in the consensus process.
Hybrid-IoT is a  permissioned blockchain system. 
This is particularly important since the sub-blockchains are based on PoW and the nature of IoT devices can easily lead to malicious cases of majority attack \cite{misbehaviorInBitcoin}. Indeed, specialized mining hardware could be easily masked as an IoT device and gain enough block mining power to control the PoW blockchain.\footnote{In the example case of Bitcoin, Raspberry Pi has 0.2 MH/s mining power, whereas a specialized mining device AntMiner S has 14 TH/s mining power.}

\textbf{Security.}
While the permissioned nature of Hybrid-IoT can mitigate the risk of longer blockchain attacks \cite{blockchainOverview}, IoT is still prone to those, since device capture and device cloning attacks are not a rare occurrence \cite{securityDistributedIoT}. Usually that would be mitigated by the difficulty of the PoW puzzle. 
Indeed, in PoW blockchains, for a fixed block generation interval, the difficulty of the PoW puzzle is set proportionally to the total mining power \cite{bitcoinNakamoto}. 
While the Hybrid-IoT PoW sub-blockchains would have PoW puzzles with low difficulty (there could be only a relatively small number of IoT devices that mine), by keeping a high block generation rate, security vulnerabilities can be prevented (see Section \ref{sec:security}).

\textbf{Anomaly resilience.}
An important issue to consider is the so called blockchain anomaly, presented in \cite{blockchainAnomaly}: "a long enough delay on the delivery of messages could lead to having the miners to seemingly agree separately on different branches containing more than \textit{k} blocks each, for any \textit{k}$\in$$\field{N}$". While this is theoretically possible, in practice, in a decentralized blockchain, like Bitcoin, it has never materialized in more than few blocks for many years.\footnote{We crawled orphan blocks through all of the Bitcoin orphan blocks presented in \url{blockchain.info/orphaned-blocks}.} Hence, in order to prevent the anomaly in Hybrid-IoT, we adopt the same degree of decentralization and scattering of mining power as in Bitcoin. That is assured by the sweet-spot guidelines and it can be further reinforced by the findings in \cite{scalingDecentralized}.

\textbf{Remediations.}
Unlike specialized PoW mining hardwares for cryptocurrencies, such as ASICs and GPUs, IoT devices have limited hardware resources and they are widely energy-constrained devices \cite{iotSurvey}. 
As such, IoT devices do not have enough hardware or energy resources to solve very complex PoW puzzles.\footnote{As of late 2017, it would require more than 1000 years for a Raspberry Pi to mine a single block in Bitcoin.} 
In Hybrid-IoT, the difficulty of the PoW puzzle is set according to the  hardware constraints of IoT devices. 
Therefore, IoT devices can still perform their application specific tasks, such as data processing, while concurrently continue to mine blocks.

\textbf{Roles of IoT devices.}
IoT devices have heterogeneous capabilities, and their roles should reflect their capabilities. Therefore, in Hybrid-IoT, we define three different roles for IoT devices as peers on the blockchain: \textit{full peer roles}; \textit{light peer roles}; and \textit{outsider roles}.

$\bullet$ \textit{Full peer role.}
 IoT devices that have enough capacity and computing power to perform complex operations, like a Raspberry Pi 3, take the \textit{full peer} role. 
They have high resources and run full-fledged operating systems like Raspbian. 
Hence, as peers on the blockchain, they mine blocks and take part in the consensus process in the  PoW sub-blockchains. In addition to that, full peer devices act as gateway devices to connect set of light peer devices to the blockchain network, referred as full peer device subnet. Hence, blocks formed by a full peer device contain its own transactions and transactions sent by its device subnet. The number of light peer devices in the full peers' device subnet is set according to its mining power to guarantee fair block generation rates.

$\bullet$ \textit{Light peer role.}
 IoT devices that have limited capabilities and computing power, such as Arduino Yun, take the \textit{light peer} role. They have basic operating systems like Alpine Linux, and can connect and participate in the blockchain by performing simple tasks, such as sending transactions. 
Light peer devices send transactions to the blockchain transaction pool and to the full peer that acts as a gateway. This allows all the full peers to be aware of all the transactions in the sub-blockchain. 
This acts as a double-check in case a full peer is subject to a malicious attack.

$\bullet$ \textit{Outsider role.}
IoT devices that have very limited capabilities by being able only to act as basic sensors, take the \textit{outsider} role. 
They are not peers on the blockchain, but they can connect to full peers for further data fusion (such as data aggregation). 
Raw data generated by an outsider is not stored the blockchain to prevent data overload.


%% file: performance.tex
\section{Performance Evaluation}\label{sec:performance}
In Hybrid-IoT, as per Section \ref{sec:architecture}, light peer devices send transactions to full peer devices, and those will include them in the newly generated blocks. Hence, full peers need to process an heavy transaction loads.
Therefore, the first performance test for Hybrid-IoT is a stress test, in which, set of light peers repeatedly sends transactions to full peers (see Section \ref{sec:perfEva1}). 
Then, we shift the focus to the different sized PoW sub-blockchains, where full peers take part in the consensus process. Sub-blockchains can be generated with different number of full peers, which affects the time required to achieve consensus and the way in which the full peer manages its resources. 
Hence, the second type of performance test is done by varying sub-blockchain sizes and measuring the time needed to achieve consensus and the full peers' resource usage (see Section \ref{sec:perfEva2}).

\subsection{Performance Evaluation \rom{1}: Stress test}\label{sec:perfEva1}
We design a DDoS attack simulation for the stress test: 20 light peers take the role of attackers; a full peer takes the role of victim; the attack is conducted for 45 minutes. All the peers are are virtualized with LXC (Linux Containers)\footnote{linuxcontainers.org} containers and have the following configurations:

$\bullet$ Full peer: Ubuntu 14.04 (Trusty) O.S; 512 MB RAM memory, 10\% of one Intel Core i7 2.70 GHz CPU; 5 mbit/s ingress and egress network interface limit; bitcoind version 14.02 Bitcoin protocol's full node. We measure  CPU utilization, memory usage, and Ethernet activity with nmon.\footnote{nmon.sourceforge.net}

$\bullet$ Light peer: Alpine Linux 3.6.0  O.S; 128 MB RAM memory, 2\% of one Intel Core i7 2.70 GHz CPU, 1 mbit/s ingress and egress network interface limit; Java SE; JRE 8 update 131 environment; Bitcoin protocol's thin client model developed with bitcoinj library. We monitor CPU usage, memory usage, and Ethernet activity with RRDtool.\footnote{oss.oetiker.ch/rrdtool}\\
We use Bitcoin regtest\footnote{bitcoin.org/en/glossary/regression-test-mode} (regression test) network to execute the stress test. The DDOS attack is executed as follows: a number of attackers, max 20, generate a load of identical and valid transactions of ca 225 bytes at varying frequency (from ca 2tx/s to ca 9tx/s); once the victim receives the load it checks the transactions validity add them to the transaction pool. A load of 108960 transactions was generated with an average of 5448 transactions per attacker. For the sake of brevity, in the figures, we show measurements only for the first 15 minutes and only for the CPU component (the other components have very similar trends).

\begin{figure*}
    \centering
   \subfloat[Perf Eva \rom{1}: CPU Utilization Light Peer ]{%
     \includegraphics[width=0.49\linewidth]{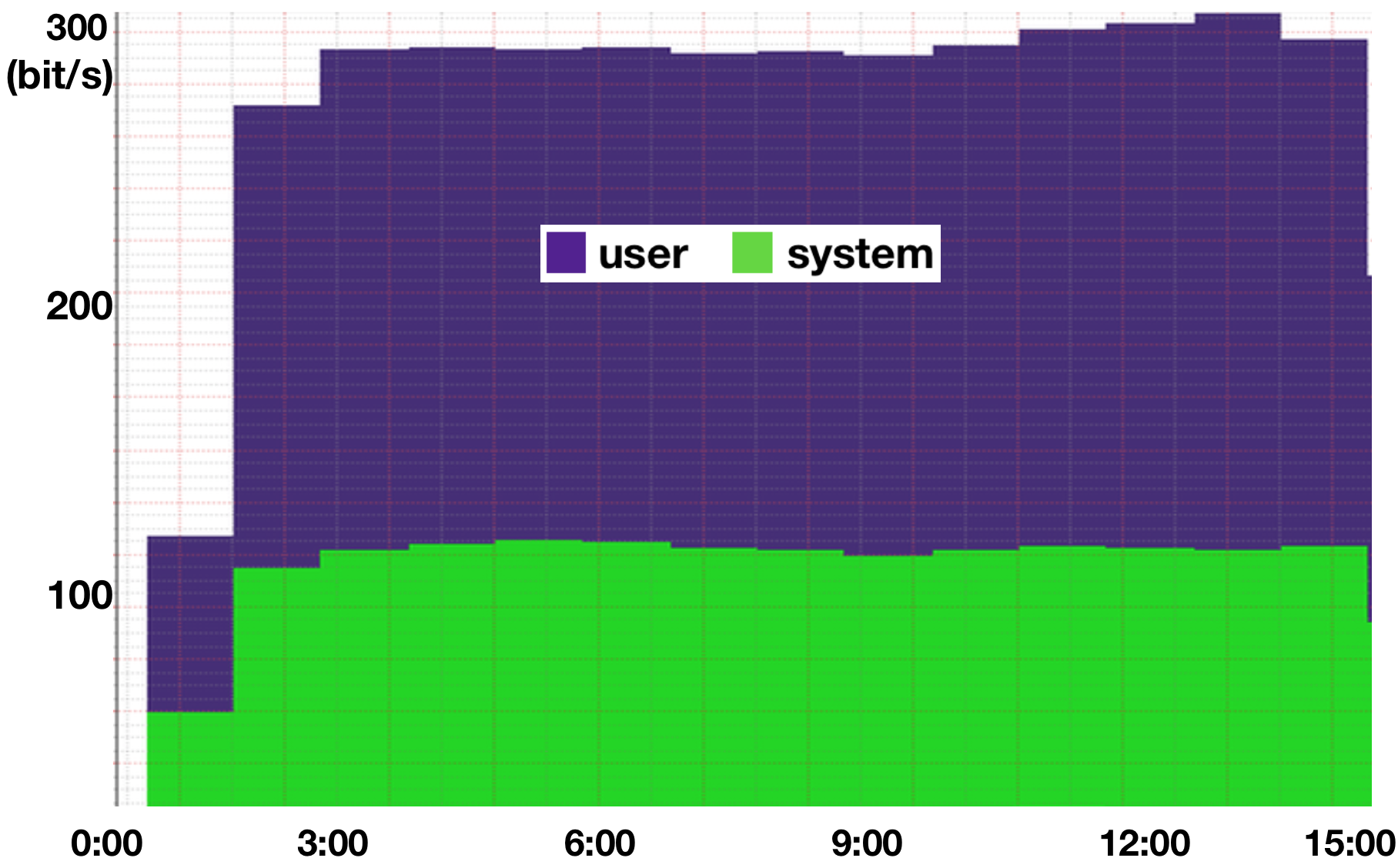}}
     \label{1a} \hfill
  \subfloat[Perf Eva \rom{1}: CPU Utilization Full Peer]{%
      \includegraphics[width=0.50\linewidth]{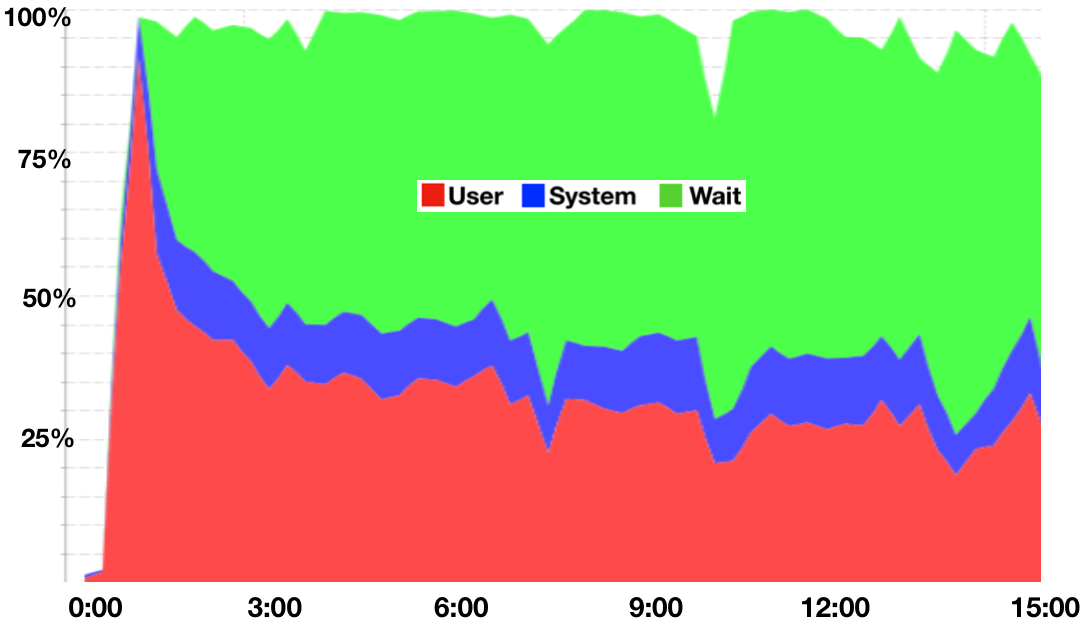}}
     \label{1b} \hfill
  \caption{CPU utilization of devices in Performance Evaluation \rom{1}}
  \label{fig:CpuUtil} 
\end{figure*}

\textbf{Victim results.} Figure \ref{fig:CpuUtil}(b)\footnote{Legend for Figure 2; user: avg CPU utilization for Bitcoin client; system: avg CPU utilization for kernel mode; wait: avg CPU utilization for I/O wait mode.} shows the CPU usage of the victim: 90\% of its CPU is exhausted by processing the attackers' load; a similar measurement and graph is observed for its Ethernet activity; memory usage is steady around 150 MBs. The victim manages to receive and process over 40 transactions per second from 20 attackers. We can conclude that the victim successfully manages to perform its blockchain duties without crashing or halting under the heavy load from the attack (here heavy is attributed to the fact that the attackers' resources are exhausted).

\textbf{Attacker results.} Figure \ref{fig:CpuUtil}(a) shows the CPU usage of one of the attackers: nearly 100\% of the attacker's CPU is exhausted (it is capable of processing ca 300 bit/s); there is an increase in memory usage from 99 MB (before starting to attack) to 124 MB (during attack), utilizing all of its memory. When one light peer takes the role of attacker, it manages a maximum of 9 transactions per second without crashing. This should help to characterize the capabilities of a light peer to generate transaction loads, regardless of the DDoS simulation performed.

\subsection{Performance evaluation \rom{2}: Sub-blockchain size}\label{sec:perfEva2}
In order to measure the sensitivity to sub-blockchain sizes, we design four sub-blockchain emulation scenarios (Emulation \rom{1},\rom{2},\rom{3} and \rom{4}) by varying the number of full peers in the sub-blockchains. 
In Emulation \rom{1} the number of full peers is 20, in Emulation \rom{2}  40, in Emulation \rom{3}  100, and in Emulation \rom{4} 200. Peers are connected to each other in a round-robin way. All the full peers are virtualized with LXC (Linux Containers)\footnote{linuxcontainers.org} containers on an IBM Power 8 server and have the following configurations:

$\bullet$ Full peer: Ubuntu 14.04 (Trusty) O.S.; 512 MB RAM memory; 5\% of a single Power8 3.5 GHz CPU; 5 mbit/s ingress and egress network interface limit; bitcoind version 14.02 Bitcoin protocol's full node. We measure  CPU utilization, memory usage, and Ethernet activity (traffic and packets) with nmon.\footnote{nmon.sourceforge.net}

We use Bitcoin regtest (regression test) network to execute the emulations. The emulations are executed as follows: in each emulation  we submit 11.000 identical transactions (225 bytes) to the network with one full peer; one peer submits to the remaining full peers the 5 blocks with 1 minute block generation interval; the full peers achieve consensus on the submitted blocks. We measure resources utilization at the last peer of the round-robin from the moment at which the submitting peer proposes the first block of the five, till the moment in which all the five blocks are recorded on the local blockchain copy of the last round-robin peer (we refer this as the consensus cycle in the rest of the paper). We note that we do not employ light peers to generate loads. This is justified by the need of measuring consensus with heavy transactions loads. 
All measurements are in Table 6 or in the text below.

\begin{table}[t]
\centering
\scriptsize
\begin{tabular}{*5{c}}      
\hline
\textbf{Metrics}   & \multicolumn{1}{c}{\textbf{20 peers}} 		& \multicolumn{1}{c}{\textbf{40 peers}}      & \multicolumn{1}{c}{\textbf{100 peers}}  & \multicolumn{1}{c}{\textbf{200 peers}} \\
\hline
Avg CPU usage         & \multicolumn{1}{c}{6.7\%}           & \multicolumn{1}{c}{5.2\%}         & \multicolumn{1}{c}{2.8\%}      & \multicolumn{1}{c}{ 2.1\%}       \\
\hline
Avg Memory usage 	     & \multicolumn{1}{c}{115 MB}          &  \multicolumn{1}{c}{109 MB}       & \multicolumn{1}{c}{109.1 MB}    & \multicolumn{1}{c}{ 108.7 MB}\\
\hline
Avg Ethernet traffic  & \multicolumn{1}{c}{7.9 KB/s}        &  \multicolumn{1}{c}{6.2 KB/s}     & \multicolumn{1}{c}{3.6 KB/s}   & \multicolumn{1}{c}{ 1.6 KB/s}\\
\hline
Avg Ethernet packets  & \multicolumn{1}{c}{9.4/s}           &  \multicolumn{1}{c}{8.5/s}        & \multicolumn{1}{c}{5/s}        & \multicolumn{1}{c}{ 3.6/s}\\
\hline
\end{tabular}
\caption{Perf Eva \rom{2}: Performance Statistics}
\end{table}\label{tab:perf2}

\textbf{Results.} 
We observe that the consensus cycle is longer with sub-blockchains with more full peers as block and transaction propagation takes longer. We show in Table 6, that on average, emulation scenarios with more full peers use less resources. This is because, with sub-blockchains with more full peers, resource utilization is averaged over longer consensus cycles.

%% file: security.tex
\section{Security Evaluation}\label{sec:security}
Despite having low difficulty puzzles, sub-blockchains can prevent security vulnerabilities with high block generation, as noted in Section \ref{sec:architecture}. We evaluate this by simulating a set of scenarios in which six sub-blockchain setups are compared. This is done with the help of the Bitcoin simulator (as in Section \ref{sec:sweetspot}) and by measuring their total work. Total work is defined as the multiplication of the number of genuine blocks by the PoW puzzle difficulty. The sub-blockchains are generated according to sweet-spot guidelines contain subgroups of IoT devices that are geographically close to each other. Hence, we generate the following scenarios:

$\bullet$ Scenario \rom{1}: 83 peers of which 6 full peers. 

$\bullet$ Scenario \rom{2}: 166 peers of which 12 full peers.

$\bullet$ Scenario \rom{3}: 250 peers of which 18 full peers.

The scenarios above are generated using the Netherlands setup. We also generate 3 more scenarios using the World setup to have a baseline:

$\bullet$ Scenario \rom{4}: 500 peers of which 36 full peers.

$\bullet$ Scenario \rom{5}: 1.000 peers of which 72 full peers.

$\bullet$ Scenario \rom{6}: 2.000 peers of which 144 full peers.

We vary the difficulty of the PoW puzzle proportionally to the number of full peers (see Table \ref{tab:secEva}). We also assume that all the peers have the same resources. In order to evaluate the sensitivity to block size we vary the block size with values 100KB, 500KB and 1MB. We configure every scenario with the shortest block generation interval, in order to be compliant with the bounds of the metrics defined in Section \ref{sec:metrics}. We present the results in Table \ref{tab:secEva}.

\begin{table}  
\centering
\begin{tabular}{*{7}{c}} 
\toprule  
\textbf{Block}  &\textbf{Simulated} &\textbf{PoW Puzzle}&\textbf{Block}    &\textbf{Genuine} &\textbf{Stale}   &\textbf{Total}  \\    
\textbf{Size}	&\textbf{Scenario}  &\textbf{Difficulty} &\textbf{Interval} &\textbf{Blocks}  &\textbf{Rate}    & \textbf{PoW} 	\\	
\hline 

		  		 & Scenario \rom{1} 	  &$\alpha$       &30s		       &198.6	 		 &1.2\%     	    & 198.6$\alpha$	\\ 
                 & Scenario \rom{2} 	  &2$\alpha$      &35s		       &162.6	 		 &1.56\%     	    & 325.2 $\alpha$	\\ 
\textbf{100 KB}  & Scenario \rom{3} 	  &3$\alpha$      &40s		       &133.2	 		 &1.7\%    	        & 299.6$\alpha$	\\ 
				 & Scenario \rom{4}    &6$\alpha$      &6m			   &15.2	 		 &1.49\%     	    & 91$\alpha$	  \\
				 & Scenario \rom{5}    &12$\alpha$     &7m			   &13.6	 		 &1.99\%  			& 163$\alpha$	  \\ 
				 & Scenario \rom{6}    &24$\alpha$     &10m			   &9.2		 		 &1.9\%     	    & 178$\alpha$	  \\
\hline 
        		 & Scenario \rom{1} 	  &$\alpha$       &50s		       &98	     		 &1.95\%   	        & 98$\alpha$	\\ 
                 & Scenario \rom{2} 	  &2$\alpha$      &55s			   &74.1	 		 &1.2\%   			& 148.2$\alpha$	  \\
\textbf{500 KB}  & Scenario \rom{3} 	  &3$\alpha$      &1m			   &90.5	 		 &1.71\%     	    & 271$\alpha$	  \\
		    	 & Scenario \rom{4}    &6$\alpha$      &10m			   &10.4	 		 &1.98\%     		& 62$\alpha$	  \\
   				 & Scenario \rom{5}    &12$\alpha$     &11m			   &9.7				 &1.89\%      		& 117$\alpha$	  \\
				 & Scenario \rom{6}    &24$\alpha$     &12m			   &8.4				 &1.96\%   			& 203$\alpha$	  \\
\hline
 			     & Scenario \rom{1} 	   &$\alpha$       &150s		   &37.5			 &1.3\%   	        &37.5$\alpha$	  \\ 
                 & Scenario \rom{2} 	   &2$\alpha$      &165s		   &34.6   	 	     &1.2\%    			&69.2 $\alpha$	  \\
\textbf{1 MB}	 & Scenario \rom{3} 	   &3$\alpha$      &3m			   &26.6   			 &0\%    			&79.8$\alpha$	  \\
		         & Scenario \rom{4}     &6$\alpha$      &10m			   &10.5			 &0.5\%      	    &63$\alpha$	  \\
				 & Scenario \rom{5}     &12$\alpha$     &12m			   &8.3				 &0\%     			&99$\alpha$	  \\
				 & Scenario \rom{6}     &24$\alpha$     &13m			   &5.9				 &1.8\%      	    &141$\alpha$	  \\
\bottomrule             
\end{tabular}
\caption{Security Experiments' Results} \label{tab:secEva} 
\end{table}

\textbf{Results.} As expected, Scenarios \rom{1}, \rom{2} and \rom{3} are able to comply with the bounds of blockchain-IoT integration metrics with shorter block generation intervals than Scenarios \rom{4}, \rom{5}, and \rom{6}, and thus they produce more genuine blocks.
This trend is more prominent with small block size settings. In fact, with 100 KB blocks, total work of Scenario \rom{1} sub-blockchain is more than the total work of Scenario \rom{6} sub-blockchain, the latter with a twenty-four times more difficult PoW puzzles than the former. Whereas, with 1 MB blocks, due to a simulated limited bandwidth (inherited from the need to replicate low bandwidth IoT), block generation intervals are longer. With 1 MB block size, total work of Scenario \rom{2} sub-blockchain is more than the Scenario \rom{4} sub-blockchain, the latter with six times more difficult PoW puzzles than the former. Hence, we observe that with smaller blocks  we can generate  sub-blockchains with less full peers without sacrificing security.
We finally observe that, even with low difficulty PoW puzzles, sub-blockchains generated according to the sweet-spot guidelines are able to have more or comparable total work than sub-blockchains with high difficulty PoW puzzles that do not adhere to those guidelines (the Netherlands scenarios have more work with easier PoW puzzles than World scenarios with more difficult PoW puzzles).

%% file: conclusions.tex
\section{Conclusions and Future Work}\label{sec:conclusions}
In this paper, we presented a novel hybrid blockchain architecture for IoT, referred to as Hybrid-IoT.
In Hybrid-IoT, subgroups of IoT devices become peers on PoW sub-blockchains, connected with a BFT inter-connector framework. In this paper, we  analyze the design of the PoW sub-blockchains.
The performance evaluation proves the validity of the PoW sub-blockchain design under the sweet-spot guidelines. Furthermore, we demonstrate that the sweet-spot guidelines also prevent security vulnerabilities.

Future work includes: analyze and stress data volumes in Hybrid-IoT; identify a BFT inter-connector framework to test the current design;
prove the correctness of Hybrid-IoT design with properly done security proofs; implement a crash fault tolerant algorithm for the light peers, to address the issue of a full peer subnet losing the connection to light peers in its subnet (due to several reasons, malicious or not); analyzing the energy footprint of PoW Hybrid-IoT; and design a PoW algorithm that is IoT energy friendly.